# Effect of Induced-Charge Electro-Kinetics on Concentration-Polarization in a Microchannel-Nafion System

Sinwook Park and Gilad Yossifon*

[1]Faculty of Mechanical Engineering, Micro- and Nanofluidics Laboratory, Technion–

Israel Institute of Technology, Technion City 32000, Israel

**Key Words**: ICEO, ACEO, Diffusion layer, Limiting Current, Concentration-polarization

PACS number(s): 47.57.jd, 82.39.Wj, 47.61.Fg, 66.10.Ed

## ABSTRACT

Induced-charge electro-osmosis (ICEO) is shown to control the length scale of the diffusion layer (DL), which in turn, affects the diffusion limited ion transport through the microchannel-Nafion membrane system. The ICEO vortices form at an interdigitated *floating* electrode array embedded within a microchannel interfacing a Nafion membrane and stir the fluid, arresting the diffusive growth of the depletion layer. By varying the spacing between the array and the membrane we are able to control the length of the DL. Activating the electrodes results in further enhancement of the fluid stirring due to the emergence of alternating-current electro-osmotic (ACEO) flow on top of the ICEO. Such a new method of controlling the DL with an electrode array is of great importance in many CP related realizations.

The passage of an electric current through a permselective membrane under an applied electric field is characterized by the formation of depleted and enriched ionic concentration regions at opposite ends of the membrane, a phenomenon termed concentration polarization (CP) [1]. In the low-voltage region, the current-voltage (I-V) behavior is approximately Ohmic until the diffusion limited current saturates when both ion concentrations are completely depleted at the surface [1]. The limiting current density scales as the inverse of the diffusion layer length and for an ideal 1D permselective membrane with negligible convection, the exact relation is $i = 2zFDc_\infty/L$. Herein, $z$ is the valency, $F$ is the Faraday constant, $D$ is the diffusion coefficient, $c_\infty$ is the bulk ionic concentration and $L$ is the diffusion layer length. Unless there is some kind of convective stirring in the system, the diffusion length spans the entire distance from the membrane interface to either the electrode or the reservoir. Since the propagation of the depletion layer results in the increase of the system resistance, its chronopotentiometric response shows a monotonic increase of the voltage. Saturation of the voltage occurs when the diffusion layer reach its finite length [2]. Hence, it is desired to be able to select a shortened diffusion length scale in applications requiring intense ion transport, e.g. electrodialysis.

A much faster saturation occurs in the case of either natural [3] or forced [4-5] convection whereby the former results in a vortex that mixes the solution while the latter is the result of the convective-diffusive boundary layer. In both cases, diffusive growth is suppressed by convection resulting in selection of a smaller diffusion layer. However, while scaling arguments [6] prove that natural convection is suppressed at the microscale, the latter complicates the system as it necessitates means of applying an external forced convection.

Electroosmotic flow of the second kind [7] is another mechanism which controls the diffusion layer length in heterogeneous systems by virtue of the resultant electro-convective stirring which can be seen in both fabricated nanochannels [8-9] and heterogeneous membranes [10]. In both these systems there is a significant tangential component of the electric field along the membrane interface which drives the extended space charge (ESC) that is induced by the normal component of the field. In contrast, for the case of homogenous permselective systems where there is a negligible tangential field component, in the absence of external stirring, electro-convection emerges only beyond a certain voltage threshold in the form of an electro-convective instability [11]. This electrokinetic instability evolves into a stationary interfacial vortex array that is found to arrest the self-similar diffusive front growth [12], which in turn specifies the overlimiting current (OLC). In the current device the microchambers are 45µm in depth, which implies that the EOF backflow mechanism will dominate the OLC behavior [13]. Thus, while the vortices due to instability may be present [14], they are not expected to contribute strongly to the OLC or suppress the depletion layer diffusive growth.

Instead, we propose a new mode of controlling the diffusion layer via the introduction of an electrode array within the microchamber and its related induced-charge electrokinetic (ICEK) effects [15], in particular, induced-charge electro-osmosis (ICEO) [16] and alternating-current electro-osmosis (ACEO) [17] occurring over floating and active electrodes, respectively. Previous studies [18], wherein similar electrode arrays were used for measurements of the local conductivity and electric fields within a microchannel, interfacing a membrane, overlooked the effect of ICEK on CP. A naive order of magnitude calculation for the induced-charge electro-osmotic (ICEO) flow on the floating electrode array, based on the solution *bulk* concentration, would erroneously seem to imply that these effects are small [19]. This scaling neglects the electric field amplification within the depletion layer [18], and since the electric field scales with the inverse of the concentration it can become very large within the depletion layer as will be shown in the following. As a result, once the depletion region extends to the electrode array, the ICEO effect becomes pronounced. The resulting ICEO induced vortices can then stir the fluid, arresting the diffusive growth of the depletion region and controlling the limiting and overlimiting currents. Activating the electrodes with alternating electric field will induce alternating-current electro-osmosis (ACEO) [17] on top of the electrodes, which act in conjunction with the ICEO produced by the external field across the membrane.

The device (Fig.1) consists of a straight Nafion membrane (3*mm* in width, 1*mm* in length and 400*nm* in depth) flanked by two polydimethysiloxane (PDMS) microchambers (3*mm* in width, 8.25*mm* in length and 45 *μm* in depth). The design is very similar to previously studied microchannel-nanochannel interface devices [14] except that instead of a fabricated nanochannel we patterned a Nafion membrane (for more details on the fabrication process – see supplementary materials [20]). An interdigitated electrode array was embedded within the microchannel at various distances (i.e. 80*μm*/0.5*mm*/2*mm*; indicated by the parameter *S* in Fig.1) from the microchannel-membrane interface. The electrode array consists of 16 pairs of electrode fingers of opposing polarity that are interlocked. The width of each electrode is $L_{electrode} = 26 \mu m$ and the gap between two adjacent electrodes is $d \sim 24$ μm . The opposite microchannel was left "bare" as a control to study the effect of the electrode array.

Two platinum wire electrodes, 0.5 mm in diameter were inserted within each reservoir (1mm in diameter with its center located ~4.75mm from the membrane interface) and connected to a voltage source (Keithley 2636) for I-V or Gamry reference 3000 in chronopotentiometry mode for V-t. The ambient temperature was maintained at 21°C, the pH=5.74 and solution conductivity $\sigma_m \approx 2.4 \mu S/cm$ were measured before each measurement. Ionic concentration profiles were extracted based on the measured fluorescent intensity of Fluorescein dye molecules (Sigma Aldrich) of 10*μM* concentration in 15μM KCl solution, normalized by the concentration intensity of the plateau away from the interface. The ICEO velocities were extracted by analyzing the dynamics of 1µm-florescent particles (Thermo Scientific) of 0.01% volumetric concentration, using IMAGE-J software and the particle tracking method. These were recorded with a spinning disc confocal system (Yokogawa CSU-X1) connected to a camera (Andor iXon3). ACEO was induced by connecting the electrode array to a function generator (Agilent 332508). For isolated ACEO frequency dispersion response, i.e. without inducing CP, smaller 500nm fluorescent particles of 0.005% volumetric concentration in the electrolyte were used.

The chronopotentiometric response of the system is depicted in Fig.2a for various applied step-wise currents. The interaction of the depleted region with the floating electrode array (i.e. "ICEO" in Fig.2a) shows a distinct change in behavior compared to the case where no such interaction exists ("Diffusion"). Specifically, the voltage drop across the system, which increases with increasing depletion region length, saturates much earlier in the presence of both the floating and activated arrays, corresponding to a much faster suppression of the diffusive layer growth (Fig.2b). The underlying mechanism behind this suppression is presumed to be the ICEO induced vortices which stir the diffusion layer (Fig.3a), arresting its diffusive growth in a manner similar to that occurring in natural or forced convective flow. The fact that this effect becomes more pronounced with increasing current also supports the dominance of these two non-linear electrokinetic mechanisms which scale quadratically with the applied voltage and at very low voltages (see I=5nA case) become negligible. Activating the electrodes and inducing ACEO on top of the ICEO, results in further suppression of the DL growth (Fig.2a, Fig.2b). Since both ACEO and ICEO are non-linear in nature, detailed analysis of their combined interaction is highly complex and is thus left as a topic for future study.

The time evolution of the depletion layer (Fig.2b) is shown in terms of the local normalized (relative to the bulk value) ionic concentration approximated from the normalized local fluorescence intensity (see movie #1 in [20]). Interestingly, instead of approaching the classical linear profile, the ionic concentration exhibits a plateau region close to the interface. Although this plateau resembles the ESC region formed at currents beyond the limiting current [21], it cannot actually be the ESC itself, which is limited to the immediate vicinity of the membrane, as the theory [21] does not account for the existence of microchannel walls and convection effects. A more viable explanation is that the plateau stems from convective mixing [18, 22] due to backflow vortices driven by both linear EOF at the microchannel walls, in particular those sections that are within the depleted region wherein the electric field is large, and EOF of the $2^{nd}$ kind stemming from the geometrical heterogeneity of the microchannel-nanoslot interface device. The Taylor-Aris-lke model of hydrodynamic dispersion has been used recently by Yaroshchuk [23] to account for the effect of mixing on the depletion layer due to backflow vortices, while the contribution of surface conduction and its coupling to bulk advection was recently addressed by Nielsen and Bruus [24].

The spatial distribution of the ICEO velocities along the electrode array in steady-state, i.e. after the CP length has been selected (~500s) (see Fig.S.1.e,f in [20]), is depicted in Fig.3b at various applied constant currents. Analyzing the colloid dynamics in the regions between the electrodes, at a fixed focal plane ~10μm above the bottom surface, using the particle tracking method, reveals both the ICEO vortex structure and the rapid decay of their intensity with increasing distance from the membrane interface. Hence, it is only the first few electrodes that predominantly contribute to the mixing. This is in agreement to the quadratic dependency of the ICEO on the local electric field as the field rapidly decays towards the edge of the depletion region.

A qualitative understanding of the effect of the depletion region on ICEO phenomena can be obtained using a simple 1D steady-state model, with the local electro-neutrality approximation (LEN) [21]. Since the electric field scales with the inverse of the concentration ($E \propto 1/c$), it increases dramatically within the depletion layer, and is at a maximum when approaching the interface. Accordingly, both the induced zeta potential ($\zeta^i \propto E \cdot L_{electrode}$) and slip velocity ($u^i \propto E^2 \cdot L_{electrode}$) increase dramatically at the electrodes within the depletion layer and are maximum at electrodes closest to the interface. However, such a simplistic model is limited and it was found (Fig.S.3 in [20]) that for the classical case of ideal permselectivity, where the concentration profile is linear in the case of limiting current, the model significantly underpredicts the slip velocity, yielding values of ($u^i \approx 0.5\,\mu m/s$, see Fig.S.3) which are markedly smaller than the experimental values ($u^i \approx 40-100\,\mu m/s$, see Fig.3). Noting that the depletion layer is not a linear distribution but rather made of a plateau-like region next to the interface with linear profile when approaching the bulk (Fig.2b), motivates the use of the modified model by Yaroshchuk [23] accounting for enhanced backflow convective mixing. For $I/I_{limiting} = 1.25$ (where $I_{limiting}$ stands for the limiting current) the non-linear analytical solution for the concentration (eq.(S.1) [20]) reveals the plateau-like region and in accordance the induced slip velocity values approach the order of those in the experiments (Fig.3c). Moreover, the predicted rapid decay of the ICEO intensity agrees with the experimental observation that it is only the first few electrodes which stir the fluid.

The effect of the spacing between the electrode array and the microchannel-Nafion interface, on the I-V response curve is depicted in Fig.4a. Here we examine both cases wherein the electrode is located at the anodic and cathodic sides of the interface. It is clearly seen that for the latter case, wherein, depletion occurs within the bare (i.e. without electrodes) microchannel, the existence of the electrodes within the opposite enriched microchannel has no influence on the I-V response and measurements from different chips collapse onto a single curve (see continuous lines in Fig.4a). However, for the former case, when the electrodes interact with the depleted region, there is a pronounced effect on the I-V response (dashed lines) at voltages beyond the Ohmic to limiting current threshold. Specifically, as electrode spacing is decreased, the OLC is increased since the interaction of the electrodes with the depleted region selects a smaller diffusion layer with larger concentration gradients leading to an increased current density. On the other hand, at large enough spacing ($S$=2mm) the selected depletion layer is too large to yield a distinct effect.

The frequency response of the ACEO was measured by activating the electrode array with no external electric field applied across the membrane. The ACEO exhibits a maximum at the RC time ($f = (\lambda_D d/D)^{-1} \approx 1.1$ kHz; wherein the Debye length $\lambda_D = \sqrt{D\varepsilon_0\varepsilon_f/\sigma_m} \sim 77$ nm, diffusion coefficient $D \sim 2 \cdot 10^{-9}$ m$^2$/s, permittivity of the vacuum $\varepsilon_0$ and dielectric constant $\varepsilon_f = 80$) (inset of Fig.4b). The low and high frequency dispersion behavior are due to the screening of the electrodes and non-sufficient time for the charging of the induced-charge layer, respectively [17]. The corresponding I-V response of the microchannel-membrane device for the case where the ACEO acts on an electrode array within the depletion layer is shown in Fig.4a. Here, the maximum effect occurs for ~1kHz, in agreement with the ACEO velocity dispersion curve. At this frequency, maximum Ohmic conductance and maximum limiting current are observed, corresponding to a minimum selected diffusion layer length (Fig.2.b and Fig.S.1 [20]). We note that these values are higher than the cases of the floating electrode array within both the anodic (ICEO) and cathodic sides (Diffusion). These differences are more pronounced at the overlimiting regime in agreement with the non-linear scaling of the combined ICEO and ACEO velocities with the electric field. As expected, in the limit of high frequency activation (~100kHz) the I-V response coincides with the floating electrode (0Vpp applied field) which is controlled by ICEO.

Thus, the presence of a floating electrode array located with a depletion layer formed due to CP across a microchannel-membrane device, is shown to dramatically suppress the diffusive growth of the depletion layer, thereby increasing the diffusion limited ion transport. The underlying mechanism is ICEO which induces vortex pairs that stir the fluid and arrest the diffusive growth. This effect is most pronounced beyond the limiting current where the existence of the depletion layer results in increased local electric field due to decreased solution conductivity. Obviously, changing the location of the electrode array affects the length of the selected DL as well as activation of the electrode which results in ACEO effect on top of the ICEO to further enhance the convective mixing. This comprehensive study of the interaction of embedded electrodes with the induced CP in microchannel-permselective medium systems, allows one to choose the electrode spacing to either enhance the mixing as a means to control the DL, or suppress it, for example, in the case where electrodes are intended for local measurements of the solution conductivity with minimal invasion. In addition the use of ACEO by activating electrodes opens new routes for on-demand control of the CP length which is of great potential in many CP based realizations.

*Acknowledgments*

We acknowledge Alicia Boymelgreen and Dr. Jarrod Schiffbauer for their valuable inputs. The fabrication of the chip was possible through the financial and technical support of the Technion RBNI (Russell Berrie Nanotechnology Institute) and MNFU (Micro Nano Fabrication Unit).

**Figure Captions**

**Fig.1:** (a) Schematics of the microchannel-Nafion nanoslot device with *S* being the spacing between the electrode array and the Nafion membrane; (b) microscopic image (top view) of the device.

**Fig.2:** (a) Chronopotentiometric response at various currents for the various cases of a floating electrode array (*S*=80μm) located at the anodic (dashed-dotted lines – "ICEO") and cathodic (continuous lines- "Diffusion") sides of the interface. The dashed lines represent the case of an electrode array activated with 5Vp-p and 1kHz ("ACEO") located at the anodic side; (b) Time evolution of the depletion layer growth (concentration normalized by its bulk value) at the anodic side of the micro-nanochannel interface (*x*=0) for *I*=15nA and electrode array (*S*=100μm) for the three cases of diffusion/ICEO/ACEO at t=250s and t=450s (see also Fig.S.1 in [20]). As seen the ICEO and the ACEO suppress the diffusive layer growth.

**Fig.3:** (a) Schematics of the ICEO vortex pairs exhibiting a decay in intensity when approaching the edge of the arrested depletion layer; (b) measured ICEO velocity distribution along the electrode array (*S*=130μm, indicated by the yellow patches) at various applied currents. The velocities were extracted by analyzing the particle dynamics in between electrodes (see inset and movie #2 in [20]); (c) theoretical prediction of the concentration (normalized by the bulk concentration) [23] and induced slip velocity (normalized by $U \equiv \left(\varepsilon_f \varepsilon_0 / \eta L\right)\left(RT/zF\right)^2$) profiles (see [20] for more details).

**Fig.4:** (a) the effect of the spacing between the electrode array and the microchannel-nanoslot interface. Both cases of electrode array located at the anodic (dashed lines) and cathodic (continuous lines) sides of the interface are examined; (b) the I-V response for varying frequencies at fixed amplitude 5Vp-p and electrode array-Nafion membrane spacing (*S*=100μm). Also the cases of no-ACEO (i.e. 0Vp-p) are depicted when the electrode array is located at both the anodic (ICEO) and cathodic sides (Diffusion). Inset showing the measured ACEO velocity frequency dispersion curve exhibiting a maximum ~1kHz.

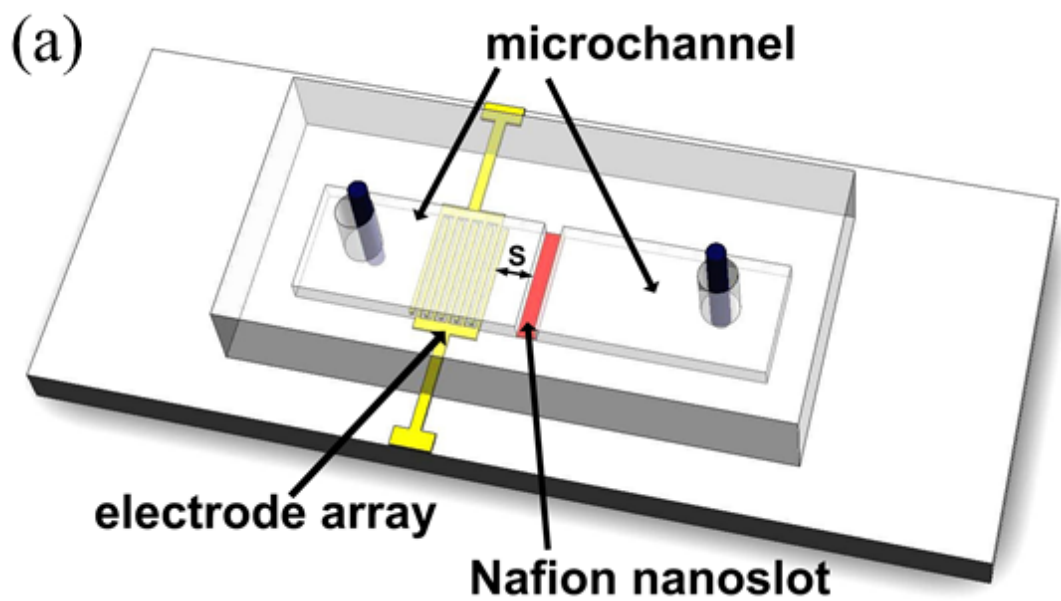

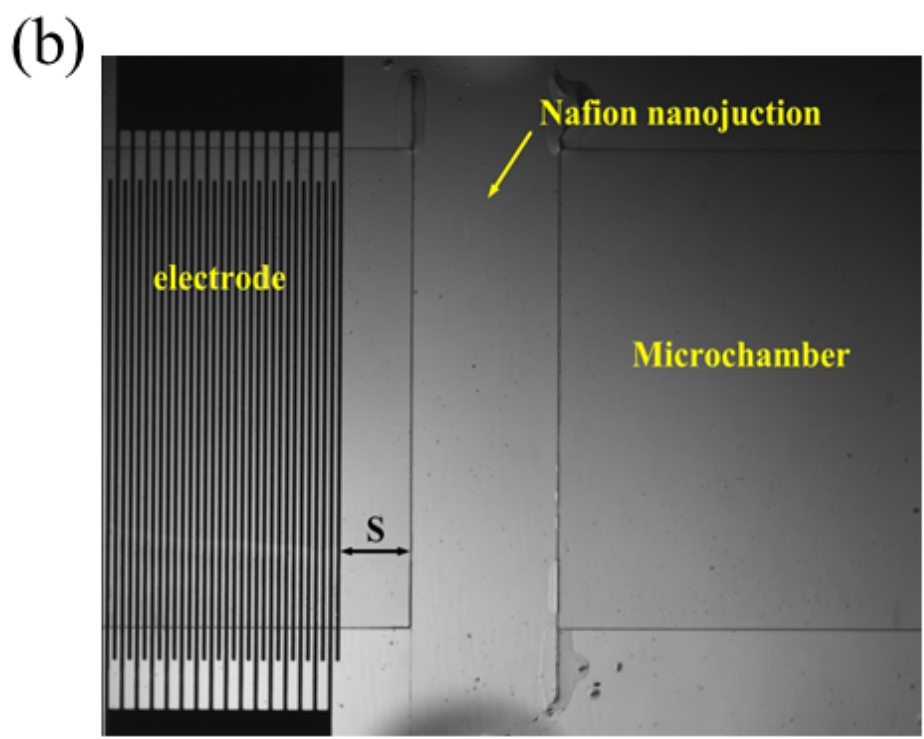

**Figure 1**

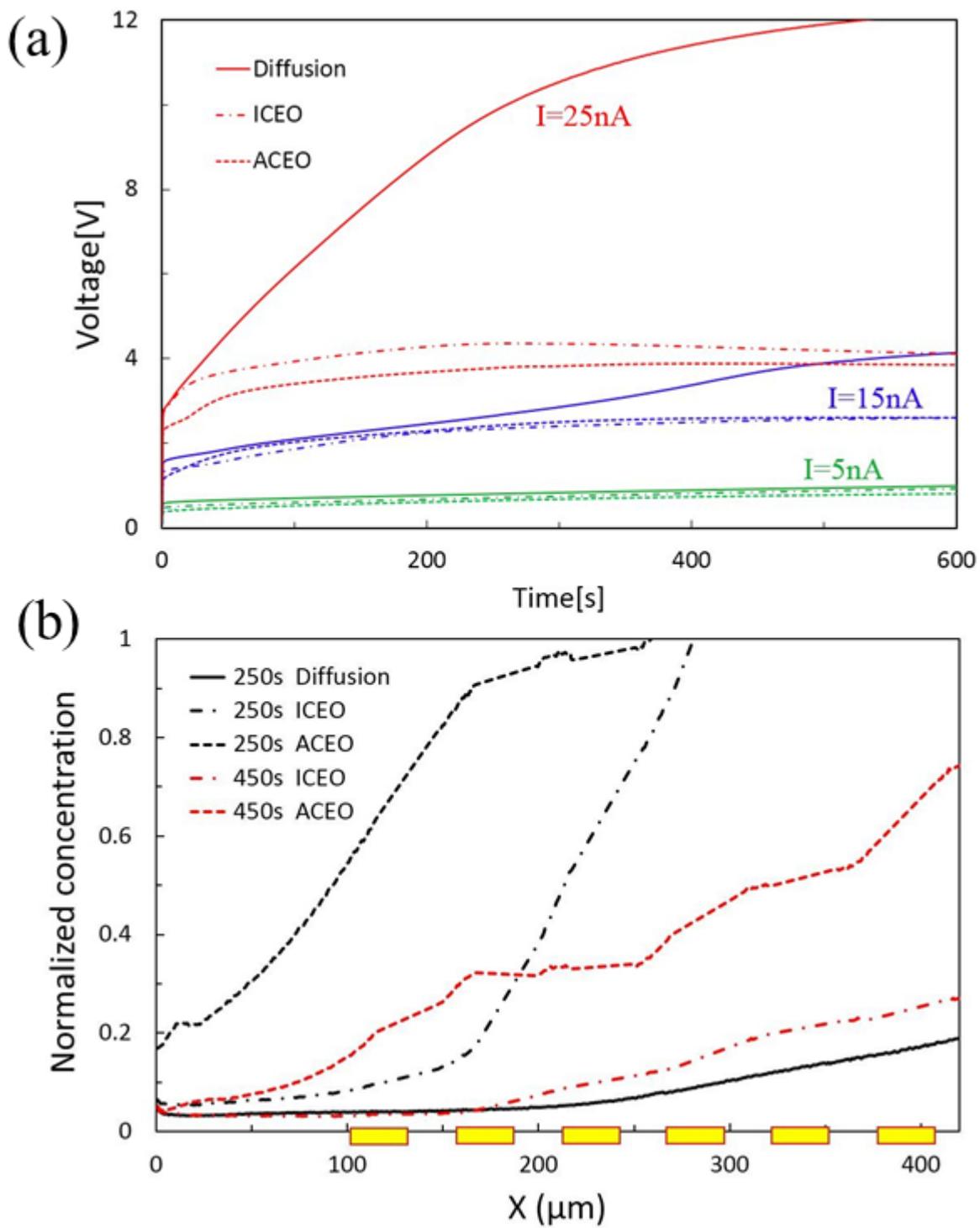

**Figure 2**

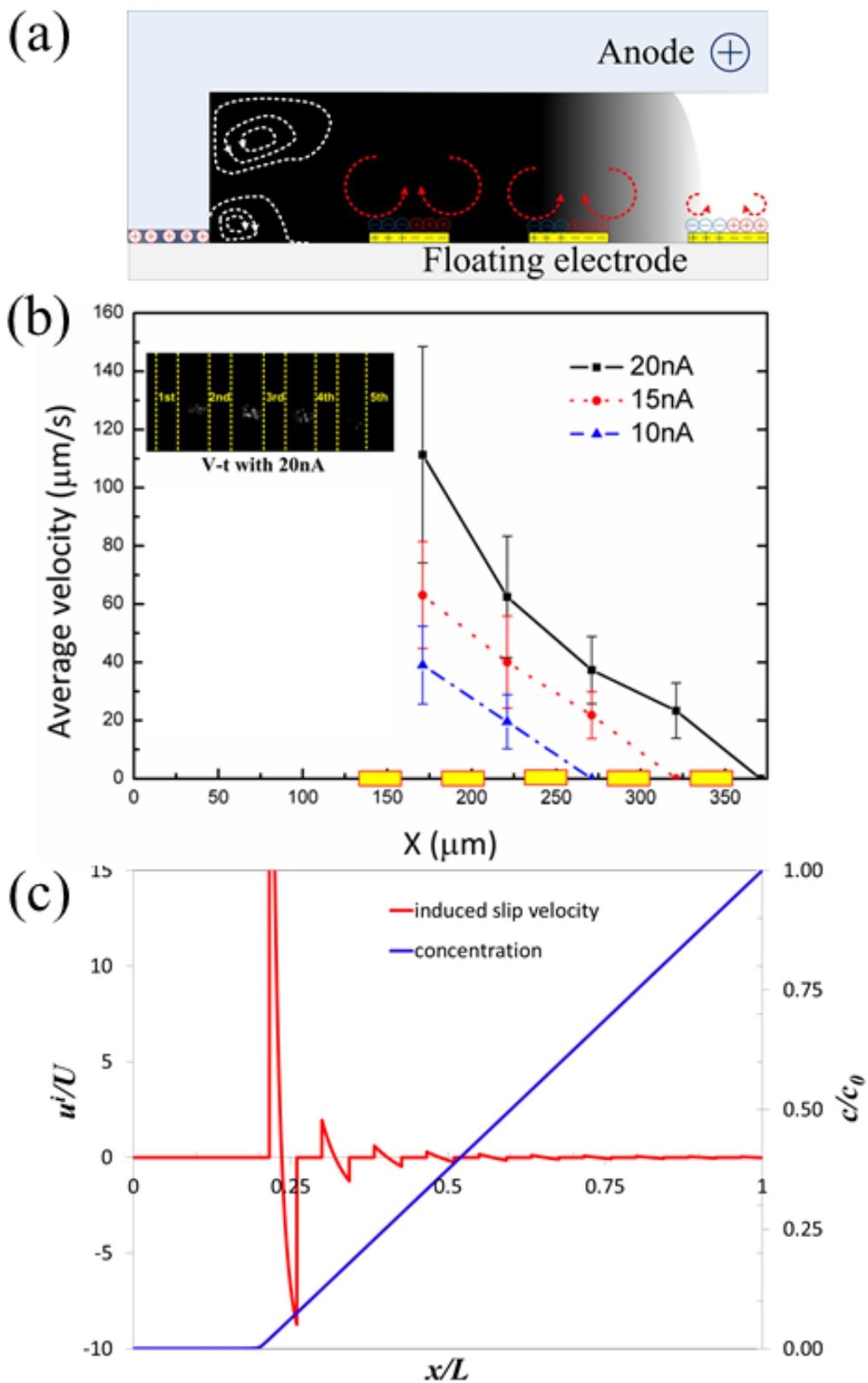

**Figure 3**

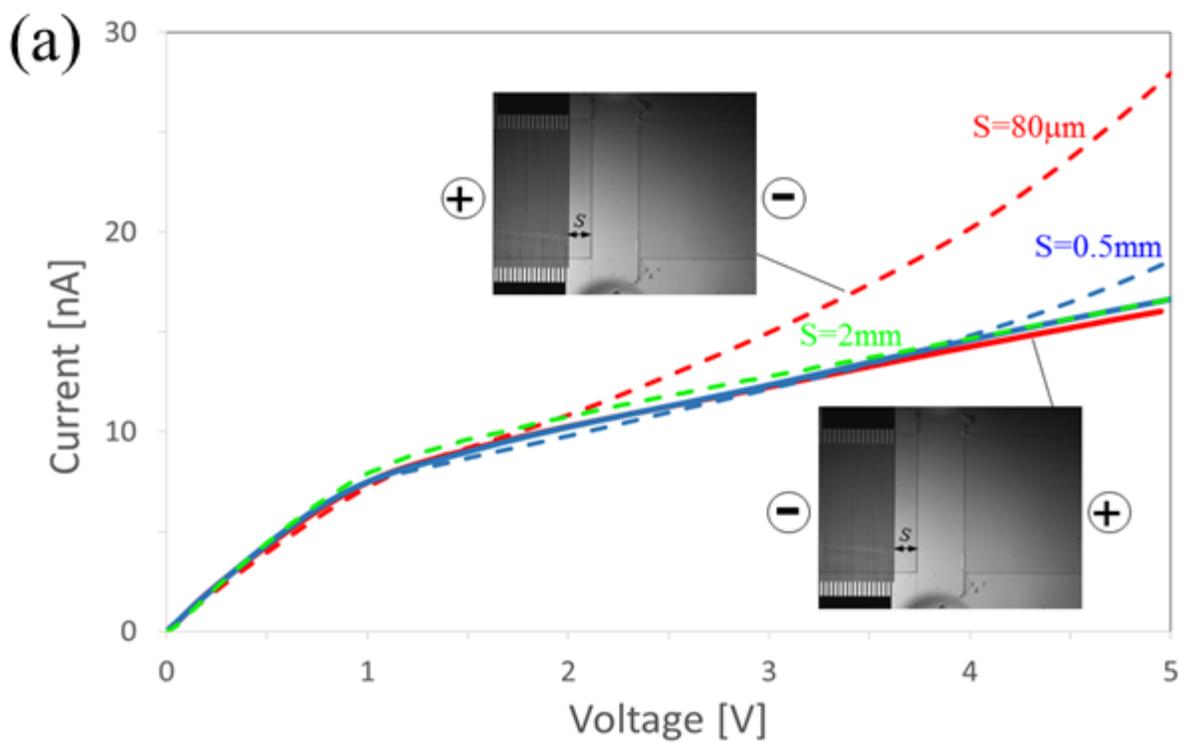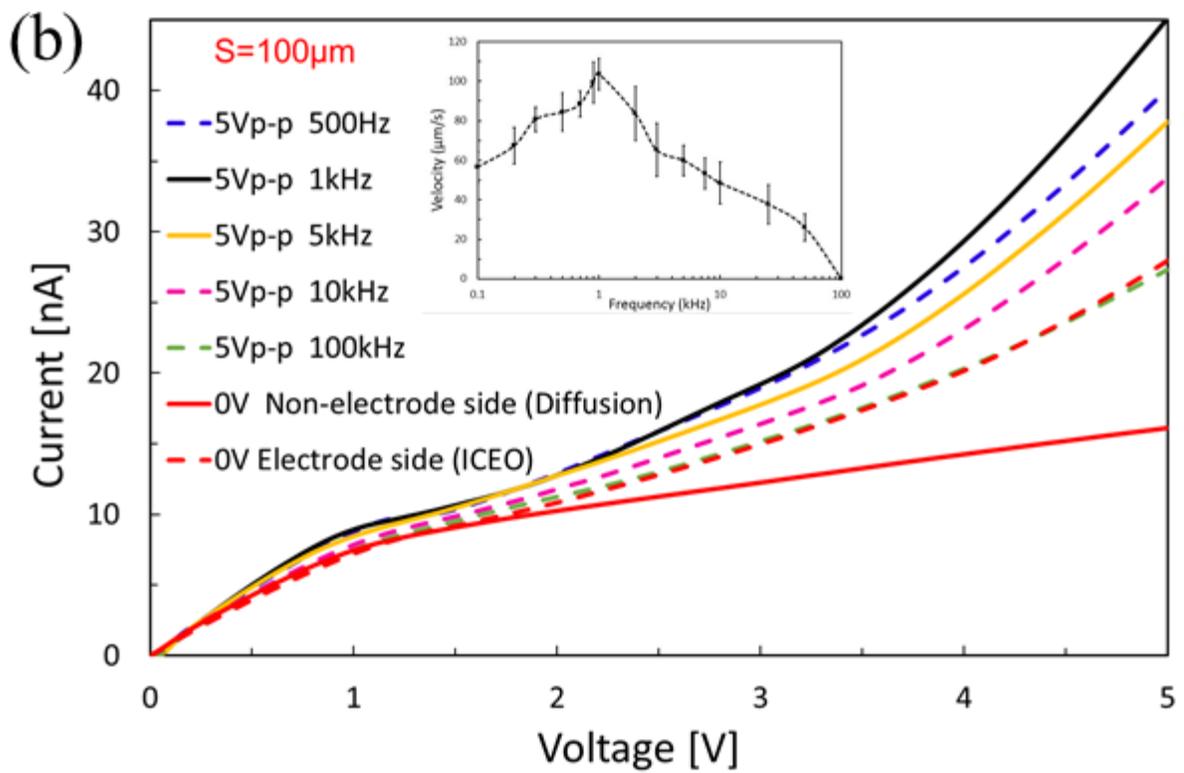

**Figure 4**